\documentclass[twocolumn,showpacs,preprintnumbers,amsmath,amssymb]{revtex4}

\usepackage{graphicx}
\usepackage{dcolumn}
\usepackage{bm}

\begin{document}


\title{Diquark Representations 
for Singly Heavy Baryons \\
with Light Staggered Quarks}

\author{Steven Gottlieb} 
\email{sg@denali.physics.indiana.edu}
\author{Heechang Na}
\email{heena@indiana.edu}
\author{Kazuhiro Nagata}
\email{knagata@indiana.edu}

\affiliation{Department of Physics, Indiana University \\
Bloomington IN 47405 USA }

\date{\today}

\begin{abstract}
In the staggered fermion formulation of lattice QCD,
we construct and classify the diquark operators 
to be embedded in the singly heavy baryons, $qqQ$. 
Along the same manner as in the staggered meson classifications given in 
\cite{Golterman},
we establish the group theoretical connections between 
continuum and lattice staggered diquark representations.

\end{abstract}

\maketitle

\section{Introduction}

There have been a number of works  addressing  
the heavy baryons from the experimental 
as well as the theoretical point of view. 
In lattice QCD, several calculations have been performed in the
quenched regime \cite{Bowler,Flynn,Mathur,Woloshyn,G-T,Khan,Chiu}. 
As for the inclusion of dynamical quarks,  
two of the authors have been studying 
the charmed and bottomed baryon mass spectrum 
using the data of $2+1$ flavors dynamical improved staggered quarks \cite{N-G}. 
Since the staggered fermion provides very fast simulations and
much less statistical errors compared to the other available frameworks, 
(see, for example, Ref \cite{MILC}),
these studies should more extensively be performed in terms of finer lattices. 
Accordingly, from the theoretical point of view, 
it is desired to establish the 
group theoretical classification
of pairs of light staggered quarks (staggered diquarks)
in order to extract the desired spin and parity state
in the singly heavy baryon analyses.
In this short report we construct all the possible time-local 
staggered diquarks embedded in singly heavy baryons.
Along the similar manner as in the meson classification \cite{Golterman},
we establish the group theoretical connection
between lattice and continuum representations w.r.t.
spin, taste and $2+1$ flavor symmetries.

\section{Staggered Diquark Representations in the Continuum}

A singly heavy baryon consists of two light quarks
(up, down or strange) and one heavy quark (charm or bottom (or top)). 
The quantum numbers of singly heavy baryons are listed in Table \ref{baryons}.
In this section, we classify the irreps
of the staggered diquarks 
w.r.t. spin, flavor and taste symmetry group 
in the continuum spacetime.
We especially take $2+1$ as the flavor symmetry group under which 
the recent dynamical simulations of staggered lattice QCD are performed.
\begingroup
\squeezetable
\begin{table}
\begin{center}
\renewcommand{\arraystretch}{1.2}
\renewcommand{\tabcolsep}{0.5pt}
\begin{tabular}{c|ccc|c}
\hline
Baryon & $J^{P}$  
& Cont. & $(SU(2)_{S}\hspace{-1pt},\hspace{-1pt}SU(2)_{I})_{Z}$ 
& ($SU(2)_{S}\hspace{-1pt},\hspace{-1pt} SU(8)_{xy}
\hspace{-1pt}, \hspace{-1pt}SU(4)_{z})_{Z}$\\
\hline
$\Lambda_{Q}$ & $\frac{1}{2}^{+}$ 
& $(ll)Q$ 
& $(\bf{1}_{A},\bf{1}_{A})_{\rm 0}$ &
$({\bf{1}_{A}}, {\bf{28}_{A}},{\mathbf{1}})_{0}$ \\
$\Xi_{Q}$ & $\frac{1}{2}^{+}$ 
&$(ls)Q$  
& $(\bf{1}_{A},\bf{2})_{\rm -1}$ & 
$({\bf{1}_{A}}, {\bf{8}},{\mathbf{4}})_{-1}$ \\
$\Sigma_{Q}^{(*)}$ & $\frac{1}{2}^{+} (\frac{3}{2}^{+})$ 
&$(ll)Q$ 
& $(\bf{3}_{S},\bf{3}_{S})_{\rm{0}}$ &
$({\bf{3}_{S}}, {\bf{36}_{S}},{\mathbf{1}})_{0}$ \\
$\Xi'^{(*)}_{Q}$ & $\frac{1}{2}^{+}(\frac{3}{2}^{+})$ 
& $(ls)Q$ 
& $(\bf{3}_{S},\bf{2})_{\rm{-1}}$ &
$({\bf{3}_{S}}, {\bf{8}},{\bf{4}})_{-1}$ \\
$\Omega^{(*)}_{Q}$ & $\frac{1}{2}^{+}(\frac{3}{2}^{+})$ 
& $(ss)Q$ 
& $(\bf{3}_{S},\bf{1})_{\rm{-2}}$ &
$({\bf{3}_{S}}, {\bf{1}},{\bf{10}_{S}})_{-2}$ \\
\hline
\end{tabular}
\caption{Quantum numbers for singly heavy baryons for
$2+1$ flavor symmetry :
$Z$ denotes strangeness.
The states with asterisks represent the spin $\frac{3}{2}$ states.
The fourth and fifth  column represent the light diquark irreps
for single taste and four tastes, respectively.}
\label{baryons}
\end{center}
\end{table}
\endgroup

Let us begin with reviewing the 
diquark irreps for mass degenerate light quarks of single taste,
namely physical valence quarks.
According to the non-relativistic $SU(6)$ quark model,
the diquarks should belong to the irreps $\bf{21}_{S}$, 
the symmetric part of $\bf{6}\otimes\bf{6}$.
It has the following decomposition into $SU(2)_{S} \times SU(3)_{F}$,
the direct product of the spin and flavor,
\begin{eqnarray}
&&SU(6) \supset SU(2)_{S} \times SU(3)_{F}  \nonumber \\
&&\bf{21}_{S} \rightarrow (\bf{3}_{S},\bf{6}_{S}) \oplus (\bf{1}_{A},\bf{3}_{A}).
\end{eqnarray} 
The labeling indicates the dimension of each irreps while
the subscripts $\bf_{S}$ and $\bf_{A}$ indicate the symmetric 
and anti-symmetric part, respectively. 
In the case of $2+1$ flavors, $SU(3)_{F}$ is decomposed into 
$SU(2)_{I}$ isospin group.
We then have
\begin{eqnarray}
&&SU(2)_{S} \times SU(3)_{F} \supset SU(2)_{S} \times SU(2)_{I}  \nonumber \\
&&(\bf{3}_{S},\bf{6}_{S}) \rightarrow
(\bf{3}_{S},\bf{3}_{S})_{\rm{0}} \oplus
(\bf{3}_{S},\bf{2})_{\rm{-1}} \oplus (\bf{3}_{S},\bf{1})_{\rm{-2}}, \\
&&(\bf{1}_{A},\bf{3}_{A}) \rightarrow
(\bf{1}_{A},\bf{1}_{A})_{\rm 0} \oplus (\bf{1}_{A},\bf{2})_{\rm -1}, 
\end{eqnarray} 
where subscripts $0,-1,-2$ denote the strangeness associated with each irrep.
Each of these irreps has one-to-one correspondence 
to the physical diquark state in singly heavy baryons,
as listed in the fourth column of Table \ref{baryons}.

As for the light staggered quarks having four tastes
with degenerate mass,
the above $SU(3)_{F}$ flavor symmetry is extended to
$SU(12)_{f}$ flavor-taste symmetry \cite{Bailey}. 
Correspondingly, the staggered diquarks belong to the symmetric irreps
of $SU(24)$ which has the following decomposition,
\begin{eqnarray}
&&SU(24) \supset SU(2)_{S} \times SU(12)_{f} \nonumber \\
&&\bf{300}_{S} \rightarrow (\bf{3}_{S},\bf{78}_{S}) \oplus (\bf{1}_{A},\bf{66}_{A}).
\label{SU(24)}
\end{eqnarray} 
For $2+1$ flavor staggered quarks, 
the $SU(12)_{f}$ flavor-taste symmetry group is broken to
$SU(8)_{xy}\times SU(4)_{z}$,
where we follow the notation given in \cite{Bailey}.
The $SU(8)_{xy}$ denotes the symmetry group for two light valence quarks
while $SU(4)_{z}$ the one for a strange valence quark. 
The decomposition of $SU(12)_{f}$ into $SU(8)_{xy}\times SU(4)_{z}$ gives,
\begingroup
\small
\begin{eqnarray}
&&SU(2)_{S}\times SU(12)_{f} \supset SU(2)_{S}\times 
SU(8)_{xy} \times SU(4)_{z} \hspace{20pt}\nonumber \\
&&({\bf{3}_{S}},\hspace{-2pt} {\bf{78}_{S}})\hspace{-1pt} \rightarrow \hspace{-1pt} 
({\bf{3}_{S}}, \hspace{-2pt}{\bf{36}_{S}},\hspace{-2pt}{\mathbf{1}})_{0} 
\hspace{-1pt}\oplus\hspace{-1pt}
({\bf{3}_{S}}, \hspace{-2pt}{\bf{8}},\hspace{-2pt}{\bf{4}})_{-1} 
\hspace{-1pt}\oplus\hspace{-1pt} 
({\bf{3}_{S}}, \hspace{-2pt}{\bf{1}},\hspace{-2pt}{\bf{10}_{S}})_{-2}, 
\label{spin_triplet} \\ 
&&({\bf{1}_{A}}, \hspace{-2pt}{\bf{66}_{A}}) \hspace{-1pt} \rightarrow\hspace{-1pt} 
({\bf{1}_{A}}, \hspace{-2pt}{\bf{28}_{A}},\hspace{-2pt}{\mathbf{1}})_{0} 
\hspace{-1pt} \oplus\hspace{-1pt}
({\bf{1}_{A}}, \hspace{-2pt}{\bf{8}},\hspace{-2pt}{\mathbf{4}})_{-1} 
\hspace{-1pt}\oplus\hspace{-1pt} 
({\bf{1}_{A}}, \hspace{-2pt}{\bf{1}},\hspace{-2pt}{\mathbf{6}_{A}})_{-2}. 
\qquad
\label{spin_singlet}
\end{eqnarray}
\endgroup
We assume the taste symmetry restores in the continuum limit and
all the four tastes become equivalent.
We then see that 
all the irreps except $({\bf{1}_{A}},{\bf{1}},{\bf{6}_{A}})_{-2}$
are to be degenerate with physical diquarks under this assumption.
We list the staggered irreps for the physical diquarks 
in the last column of Table \ref{baryons}.
The strangeness $-2$ spin singlet diquark state
$({\bf{1}_{A}},{\bf{1}},{\bf{6}_{A}})_{-2}$ in (\ref{spin_singlet})
is obviously peculiar to the staggered representation
and it does not correspond to any physical state in the continuum limit.

In order to make contact with the lattice symmetry group,
we further need to categorize the physical states
w.r.t. $SU(2)_{S} 
\times SU(4)_{T}$, the spin and taste symmetries,
since the lattice rotation group for the staggered fermion
is embedded in the diagonal subgroup of 
spacetime and taste rotation \cite{Doel-Smit,G-S_self_energy}.
We first decompose the entire flavor-taste symmetry group
$SU(12)_{f}$ into the direct product of $SU(3)_{F}$  and $SU(4)_{T}$,
\begin{eqnarray}
SU(12)_{f} &\supset& SU(3)_{F} \times SU(4)_{T}  \nonumber \\[2pt] 
\bf{78}_{S} &\rightarrow& (\bf{6}_{S},\bf{10}_{S}) \oplus (\bf{3}_{A},\bf{6}_{A}), 
\label{decomp1}\\[2pt]
\bf{66}_{A} &\rightarrow& (\bf{6}_{S},\bf{6}_{A}) \oplus (\bf{3}_{A},\bf{10}_{S}).
\label{decomp2}
\end{eqnarray}
and successively into the $2+1$ flavors. 
We then have
\begingroup
\small
\begin{eqnarray}
&&SU(2)_{S}\hspace{-2pt}\times \hspace{-2pt}SU(8)_{xy}
\hspace{-2pt}\times \hspace{-2pt}
SU(4)_{z} 
\hspace{-2pt}\supset\hspace{-2pt}
SU(2)_{S} \hspace{-2pt}\times \hspace{-2pt} 
SU(2)_{I} \hspace{-2pt}\times \hspace{-2pt}SU(4)_{T} \nonumber \\
&&\Sigma_{Q}^{(*)}
: 
(\bf{3}_{S}, \hspace{-1pt}\bf{36}_{S},\hspace{-1pt}\bf{1})_{\rm 0} 
\rightarrow 
(\bf{3}_{S},\hspace{-1pt}\bf{3}_{S},\hspace{-1pt}\bf{10}_{S})_{\rm 0} 
\hspace{-1pt}\oplus\hspace{-1pt}
(\bf{3}_{S},\hspace{-1pt}\bf{1}_{A},\hspace{-1pt}\bf{6}_{A})_{\rm 0}, 
\label{Sigma_2}\\
&&\Xi'^{(*)}_{Q}
:  
(\bf{3}_{S},\hspace{-1pt}\bf{8},\hspace{-1pt}\bf{4})_{\rm -1}
\rightarrow
(\bf{3}_{S},\hspace{-1pt}\bf{2},\hspace{-1pt}\bf{10}_{S})_{\rm -1} 
\hspace{-1pt}\oplus\hspace{-1pt}
(\bf{3}_{S},\hspace{-1pt}\bf{2},\hspace{-1pt}\bf{6}_{A})_{\rm -1}, \\
&&\Omega_{Q}^{(*)}
: (\bf{3}_{S},\bf{1},\bf{10}_{S})_{\rm -2} 
\rightarrow
(\bf{3}_{S},\hspace{-1pt}\bf{1},\hspace{-1pt}\bf{10}_{S})_{\rm -2}, \\
&&\Lambda_{Q} : 
(\bf{1}_{A},\hspace{-1pt}\bf{28}_{A},\hspace{-1pt}\bf{1})_{\rm 0} 
\rightarrow 
(\bf{1}_{A},\hspace{-1pt}\bf{1}_{A},\hspace{-1pt}\bf{10}_{S})_{\rm 0} 
\hspace{-1pt}\oplus\hspace{-1pt}
(\bf{1}_{A},\hspace{-1pt}\bf{3}_{S},\hspace{-1pt}\bf{6}_{A})_{\rm 0},\\
&&\Xi_{Q} : 
(\bf{1}_{A},\hspace{-1pt}\bf{8},\hspace{-1pt}\bf{4})_{\rm -1} 
\rightarrow 
(\bf{1}_{A},\hspace{-1pt}\bf{2},\hspace{-1pt}\bf{10}_{S})_{\rm -1} 
\hspace{-1pt}\oplus\hspace{-1pt}
(\bf{1}_{A},\hspace{-1pt}\bf{2},\hspace{-1pt}\bf{6}_{A})_{\rm -1}. \label{Xi_2}
\end{eqnarray}
\endgroup
The main goal of this report is to construct the lattice staggered diquark
operators categorized into the physical irreps given in 
the right hand sides of
(\ref{Sigma_2})-(\ref{Xi_2}).

\section{Staggered Diquark Representations on the Lattice}

\begingroup
\begin{table*}[t]
\renewcommand{\arraystretch}{0.8}
\renewcommand{\tabcolsep}{8pt}
\begin{scriptsize}
\begin{tabular}{ccccccccc}
\hline
class & No. & operator & $GTS$ ($\overline{\mathbf{r}}^{\sigma_{s}\sigma}$) &
$\sigma_{4}$ & $\Gamma_{S}\otimes \Gamma_{T}$  
& $J^{P}$  & order & $(SU(2)_{S},SU(4)_{T})$\\ \hline
$0$ & $1$ & $\chi\chi$ & $\mathbf{1}^{++}$ & $+$ & $\gamma_{5}\otimes\gamma_{5}$ 
& $0^{+}$ &   $1$ & $(\bf{1}_{A},\bf{6}_{A})$\\
& & &  & $-$ & $\gamma_{4}\otimes\gamma_{4}$ 
& $0^{-}$ & $p/E$\\ 
& $2$ & $\eta_{4}\zeta_{4}\chi\chi$ 
& $\mathbf{1}^{+-}$ & $+$ & $\gamma_{4}\gamma_{5}\otimes\gamma_{4}\gamma_{5}$ 
 & $0^{+}$ & $1$ & $(\bf{1}_{A},\bf{6}_{A})$\\
&&&  & $-$ & $1\otimes1$ 
& $0^{-}$ & $p/E$\\ 
& $3$ & $\eta_{k}\epsilon\zeta_{k}\chi\chi$ 
& $\mathbf{3''''}^{+-}$ & $+$ & $\gamma_{k}\otimes\gamma_{k}$ 
& $1^{+}$ &  $1$ 
& $(\bf{3}_{S},\bf{10}_{S})$\\
&&&  & $-$ & $\gamma_{l}\gamma_{m}\otimes\gamma_{l}\gamma_{m}$ 
&$1^{-}$ &  $p/E$\\ 
&$4$ &$\eta_{4}\zeta_{4}\eta_{k}\epsilon\zeta_{k}\chi\chi$ 
& $\mathbf{3''''}^{++}$ & $+$ & $\gamma_{k}\gamma_{4}\otimes\gamma_{k}\gamma_{4}$ 
&$1^{+}$ &  $1$
& $(\bf{3}_{S},\bf{10}_{S})$\\
&&&  & $-$ & $\gamma_{k}\gamma_{5}\otimes\gamma_{k}\gamma_{5}$ 
& $1^{-}$ & $p/E$\\[10pt] 
$1$ & $5$ &
$\chi\eta_{k}D_{k}\chi$ 
& $\mathbf{3}^{-+}$ & $+$ & $\gamma_{k}\gamma_{5}\otimes\gamma_{5}$ 
& $1^{-}$ & $p/E$ \\
&&&  & $-$ & $\gamma_{k}\gamma_{4}\otimes\gamma_{4}$ 
&$1^{+}$ & $1$
& $(\bf{3}_{S},\bf{10}_{S})$\\ 
&$6$ &$\eta_{4}\zeta_{4}\chi\eta_{k}D_{k}\chi$ 
& $\mathbf{3}^{--}$ & $+$ & $\gamma_{l}\gamma_{m}\otimes\gamma_{4}\gamma_{5}$ 
& $1^{-}$ &  $p/E$\\
&&&  & $-$ & $\gamma_{k}\otimes 1$ 
&$1^{+}$ &  $1$
& $(\bf{3}_{S},\bf{6}_{A})$ \\ 
&$7$ &$\chi\epsilon\zeta_{k}D_{k}\chi$ 
& $\mathbf{3}''^{--}$ & $+$ & $1\otimes\gamma_{k}$ 
& $0^{-}$ &  $p/E$\\
&&&  & $-$ & $\gamma_{4}\gamma_{5}\otimes \gamma_{l}\gamma_{m}$ 
& $0^{+}$ &  $1$
& $(\bf{1}_{A},\bf{10}_{S})$\\ 
&$8$&$\eta_{4}\zeta_{4}\chi\epsilon\zeta_{k}D_{k}\chi$ 
& $\mathbf{3}''^{-+}$ & $+$ & $\gamma_{4}\otimes\gamma_{k}\gamma_{4}$ 
&$0^{-}$ & $p/E$\\
&&&  & $-$ & $\gamma_{5}\otimes \gamma_{k}\gamma_{5}$ 
& $0^{+}$ & $1$ & $(\bf{1}_{A},\bf{6}_{A})$\\ 
&$9$ &$\eta_{k}\epsilon\zeta_{k}\chi\eta_{l}D_{l}\chi$ 
& $\mathbf{6}^{--}$ & $+$ & $\gamma_{k}\gamma_{l}\otimes\gamma_{k}$ 
& $1^{-}$ &  $p/E$\\
&&&  & $-$ & $\gamma_{m}\otimes \gamma_{l}\gamma_{m}$ 
& $1^{+}$ & $1$
& $(\bf{3}_{S},\bf{10}_{S})$\\ 
&$10$ & $\eta_{4}\zeta_{4}\eta_{k}\epsilon\zeta_{k}\chi\eta_{l}D_{l}\chi$ 
& $\mathbf{6}^{-+}$ & $+$ & $\gamma_{m}\gamma_{5}\otimes\gamma_{k}\gamma_{4}$ 
& $1^{-}$ &  $p/E$\\
&&&  & $-$ & $\gamma_{m}\gamma_{4}\otimes \gamma_{k}\gamma_{5}$ 
& $1^{+}$ &  $1$ & $(\bf{3}_{S},\bf{6}_{A})$  \\ \hline
\end{tabular}
\end{scriptsize}
\caption{$GTS$ irrep., $\sigma_{4}$, $\Gamma_{S}\otimes \Gamma_{T}$
and continuum states for staggered diquark operators up to class 1.
($k,l,m = 1\sim 3,\ k\neq l \neq m \neq k$).
The summation over $\bf{x}$, flavor and color indices are omitted.}
\label{diquark_operators}
\end{table*}
\endgroup

\begin{table*}
\begin{center}
\renewcommand{\arraystretch}{0.8}
\renewcommand{\tabcolsep}{5pt}
\begin{scriptsize}
\begin{tabular}{ccccccccc}
\hline
class & No. & operator & $GTS$ ($\overline{\mathbf{r}}^{\sigma_{s}\sigma}$) &
$\sigma_{4}$ & $\Gamma_{S}\otimes \Gamma_{T}$  
& $J^{P}$ &  order & $(SU(2)_{S},SU(4)_{T})$\\ \hline
$2$ &$11$& $\chi\eta_{k}D_{k}\{\eta_{l}D_{l}\chi\}$ 
& $\mathbf{3}^{++}$ & $+$ & $\gamma_{m}\gamma_{4}\otimes\gamma_{5}$ 
& $1^{+}$ &  $1 $ & $(\bf{3}_{S},\bf{6}_{A})$\\
& & && $-$ & $\gamma_{m}\gamma_{5}\otimes\gamma_{4}$
& $1^{-}$ &  $p/E$ \\
 & $12$&$\eta_{4}\zeta_{4}\chi\eta_{k}D_{k}\{\eta_{l}D_{l}\chi\}$ 
& $\mathbf{3}^{+-}$ & $+$ & $\gamma_{m}\otimes\gamma_{4}\gamma_{5}$ 
& $1^{+}$ &  $1$ & $(\bf{3}_{S},\bf{6}_{A})$\\
& && & $-$ & $\gamma_{k}\gamma_{l}\otimes 1$
& $1^{-}$ &  $p/E$\\
 & $13$ &$\chi\zeta_{k}D_{k}\{\zeta_{l}D_{l}\chi\}$ 
& $\mathbf{3}''^{++}$ & $+$ & $\gamma_{5}\otimes\gamma_{m}\gamma_{4}$ 
& $0^{+}$ &  $1$
& $(\bf{1}_{A},\bf{10}_{S})$ \\
& & & &$-$ & $\gamma_{4}\otimes \gamma_{m}\gamma_{5}$
& $0^{-}$ &  $p/E$\\
 & $14$ &$\eta_{4}\zeta_{4}\chi\zeta_{k}D_{k}\{\zeta_{l}D_{l}\chi\}$ 
& $\mathbf{3}''^{+-}$ & $+$ & $\gamma_{4}\gamma_{5}\otimes\gamma_{m}$ 
& $0^{+}$ &  $1$
& $(\mathbf{1}_{A},\mathbf{10}_{S})$\\
& & && $-$ & $1 \otimes \gamma_{k}\gamma_{l}$
& $0^{-}$ &  $p/E$\\
 & $15$ &$\eta_{m}\zeta_{m}\chi\eta_{k}D_{k}\{\zeta_{l}D_{l}\chi\}$ 
& $\mathbf{6}^{++}$ & $+$ & $\gamma_{l}\gamma_{4}\otimes\gamma_{k}\gamma_{4}$ 
& $1^{+}$ &  $1$ 
& $(\bf{3}_{S},\bf{10}_{S})$\\
& & && $-$ & $\gamma_{l}\gamma_{5} \otimes \gamma_{k}\gamma_{5}$
& $1^{-}$ &  $p/E$\\
 & $16$ &$\eta_{4}\zeta_{4}\eta_{m}\zeta_{m}\chi\eta_{k}D_{k}\{\zeta_{l}D_{l}\chi\}$ 
& $\mathbf{6}^{+-}$ & $+$ & $\gamma_{l}\otimes\gamma_{k}$ 
& $1^{+}$ &  $1$
& $(\bf{3}_{S},\bf{10}_{S})$\\
& & & &$-$ & $\gamma_{k}\gamma_{m} \otimes \gamma_{l}\gamma_{m}$
& $1^{-}$ &  $p/E$\\[10pt]
$3$ &$17$ & $\chi\eta_{1}D_{1}\{\eta_{2}D_{2}\{\eta_{3}D_{3}\chi\}\}$ 
& $\mathbf{1}^{-+}$ & $+$ & $\gamma_{4}\otimes\gamma_{5}$ 
& $0^{-}$ &  $p/E$\\
& & & & $-$ & $\gamma_{5} \otimes \gamma_{4}$
& $0^{+}$ &  $1$
&  $(\bf{1}_{A},\bf{10}_{S})$\\
 & $18$ &$\eta_{4}\zeta_{4}\chi\eta_{1}D_{1}\{\eta_{2}D_{2}\{\eta_{3}D_{3}\chi\}\}$ 
& $\mathbf{1}^{--}$ & $+$ & $ 1\otimes\gamma_{4}\gamma_{5}$ 
& $0^{-}$ &  $p/E$\\
& & & &$-$ & $\gamma_{4}\gamma_{5} \otimes 1$
& $0^{+}$ &  $1$ & $(\bf{1}_{A},\bf{6}_{A})$\\
 & $19$ &$\eta_{k}\epsilon\zeta_{k}\chi\eta_{1}D_{1}
\{\eta_{2}D_{2}\{\eta_{3}D_{3}\chi\}\}$ 
& $\mathbf{3}''''^{--}$ & $+$ & $ \gamma_{l}\gamma_{m}\otimes \gamma_{k}$ 
& $1^{-}$ &  $p/E$\\
& & & &$-$ & $\gamma_{k}\otimes \gamma_{l}\gamma_{m}$
& $1^{+}$ & $1$
& $(\bf{3}_{S},\bf{10}_{S})$\\
 & $20$ &$\eta_{4}\zeta_{4}\eta_{k}\epsilon\zeta_{k}
\chi\eta_{1}D_{1}\{\eta_{2}D_{2}\{\eta_{3}D_{3}\chi\}\}$ 
& $\mathbf{3}''''^{-+}$ & $+$ & $ \gamma_{k}\gamma_{5}\otimes \gamma_{k}\gamma_{4}$ 
& $1^{-}$ &  $p/E$\\
& & && $-$ & $\gamma_{k}\gamma_{4}\otimes \gamma_{k}\gamma_{5}$
& $1^{+}$ &  $1$ & $(\bf{3}_{S},\bf{6}_{A})$ \\
\hline
\end{tabular}
\end{scriptsize}
\caption{$GTS$ irrep., $\sigma_{4}$, $\Gamma_{S}\otimes \Gamma_{T}$
and continuum states for staggered diquark operators class 2 and 3.
($k,l,m = 1\sim 3,\ k\neq l \neq m \neq k$).
The summation over $\bf{x}$, flavor and color indices are omitted.}
\label{diquark_operators2}
\end{center}
\end{table*}


The symmetry group of staggered fermion action
on the Euclidean lattice was first elaborated in \cite{Doel-Smit,G-S_self_energy}
and successively applied to classifying staggered 
baryons as well as mesons \cite{G-S,Golterman,Smit}.
The important symmetries of staggered fermions 
we need to look at in our study are the
$90^{\circ}$ 
rotations $R^{(\rho\sigma)}$, the shift transformations $S_{\mu}$, 
and the space inversion $I_{s}$.
Since the shift operations $S_{\mu}$ contain
taste matrices,
pure translations $T_{\mu}$ may be represented by the square of $S_{\mu}$,
$T_{\mu}=S_{\mu}^{2}$.
Discrete taste transformations in Hilbert space are readily defined
by 
$\Xi_{\mu}\equiv S_{\mu}T^{-\frac{1}{2}}_{\mu}. \label{Xi_mu} $
The $\Xi_{\mu}$ generate 32 element Clifford group which 
is isomorphic to the discrete subgroup of $SU(4)_{T}$ in the continuum spacetime.
The representations of $\Xi_{\mu}$, $D_{q}(\Xi_{\mu})$, 
for a given quark number $q$, obey 
\begin{eqnarray}
D_{q}(\Xi_{\mu})D_{q}(\Xi_{\nu})
=e^{i\pi q}D_{q}(\Xi_{\nu})D_{q}(\Xi_{\mu}). 
\end{eqnarray}
Since the space inversion contains a taste transformation $\Xi_{4}$,
the parity should be defined by 
\begin{eqnarray}
 P=\Xi_{4}I_{s}. \label{def_parity}
\end{eqnarray}
Note that the parity is non-locally defined in time direction 
since $\Xi_{4}$ is non-local in time.
For the purpose of spectroscopy, 
we are particularly interested in a symmetry group generated by 
the transformations which are local in time and
commuting with $T_{4}$.
Such a group is called geometrical time slice group ($GTS$)
which is given by
\begin{eqnarray}
GTS &=& G(R^{(kl)},\Xi_{m},I_{s})
\end{eqnarray}
where $k,l, m = 1\sim 3$ \cite{G-S,Golterman,Smit}.
The defining representation of $GTS$ is 
given by the staggered quark fields
projected on zero spatial momentum.
It is an eight dimensional representation denoted as $\bf{8}$.
The anti-staggered quark fields 
also belong to the representation $\bf{8}$.
The $GTS$ representation of
staggered diquark is accordingly expressed by $\mathbf{8}\times\mathbf{8}$.
The decomposition of $\mathbf{8}\times\mathbf{8}$
into the bosonic irreps is given in \cite{Golterman},
\begin{eqnarray}
\mathbf{8}\times\mathbf{8} = \hspace{-4pt} 
\sum_{\sigma_{s}, \sigma}
\{
\mathbf{1}^{\sigma_{s}\sigma}
\hspace{-1pt}+\hspace{-1pt}
\mathbf{3}^{\sigma_{s}\sigma}
\hspace{-1pt}+\hspace{-1pt}
\mathbf{3''}^{\sigma_{s}\sigma}
\hspace{-1pt}+\hspace{-1pt}
\mathbf{3''''}^{\sigma_{s}\sigma}
\hspace{-1pt}+\hspace{-1pt}
\mathbf{6}^{\sigma_{s}\sigma}
\} \hspace{5pt} \label{8times8}
\end{eqnarray}
where
$\bf{1}$, $\bf{3}$, $\bf{3''}$, $\bf{3''''}$
and $\bf{6}$ are representing the bosonic representations
of $GTS$.
The $\sigma_{s}$ and 
$\sigma$
are denoting the eigenvalue of $I_{s}$
and $D(\Xi_{1}\Xi_{2}\Xi_{3})$, respectively. 
In Eq. (\ref{8times8}),
the $\sigma_{s}$ and $\sigma$ are summed over $\sigma_{s}=\pm 1$
and $\sigma =\pm 1$
respectively.

The irreducibly transforming diquark operators  
are listed in Table \ref{diquark_operators} and \ref{diquark_operators2}.
As in the meson case, all the irreps are categorized 
into four classes from 0 to 3,
depending on how far the two staggered quarks are displaced each other.
The third column of the tables gives the operator form
of the diquarks.
The fourth column
gives the corresponding $GTS$ irreps.  
The $\eta_{\mu}$ and $\zeta_{\mu}$ denote the sign
factors defined by
$\eta_{\mu}(x)=(-1)^{x_{1}+\cdots +x_{\mu-1}}$ 
and $\zeta_{\mu}(x)=(-1)^{x_{\mu +1}+\cdots +x_{4}}$, respectively,
while $\epsilon$ is defined as $\epsilon(x)=(-1)^{x_{1}+x_{2}+x_{3}+x_{4}}$.  
The $D_{k}$ represents the symmetric shift operators defined by
$D_{k}\phi({\bf x}) = \frac{1}{2}[\phi({\bf x} +{\bf a}_{k})
+\phi({\bf x} -{\bf a}_{k})]$.
For notational simplicity, the sum over $\mathbf{x}$,
the color and flavor indices are suppressed. 
For example, $\chi \eta_{k}D_{k} \chi$
stands for
$
\sum_{\mathbf{x}}\chi^{a}_{f_{1}}({\bf x},t) 
\eta_{k}(x) D_{k} \chi^{b}_{f_{2}}({\bf x},t)
$.
As far as the lattice symmetry group $GTS$ is concerned, 
each diquark operator is formally corresponding to the meson operator
given in \cite{Golterman} through replacing the leftmost $\chi$ by $\overline{\chi}$.
This is because the staggered quark and anti-quark 
belong to the same $GTS$ irrep for each color and flavor.
The $\sigma_{4}$ in the fifth column denotes the eigenvalue of $X_{4}$ 
with which the parity 
is given by $P=\sigma_{s}\sigma_{4}$.
The sixth column gives
the spin and taste matrices 
$\Gamma_{S}\otimes\Gamma_{T}$
which come into the diquark operators in the spin-taste basis,
$
\psi^{T}(C\Gamma_{S}\otimes (\Gamma_{T}C^{-1})^{T})\psi,
$
where the superscript $T$ denotes transpose
and $C$ denotes the charge conjugation matrix.
The presence of $C$ and $C^{-1}$
ensures
the covariant properties under the spin and taste 
rotations in the continuum limit.
Notice that
the assignment of $\Gamma_{S}\otimes \Gamma_{T}$ for 
each $GTS$ irrep is systematically different from the meson case, 
where the operators are given by 
$\overline{\psi}(\Gamma_{S}\otimes(\Gamma_{T})^{T})\psi$
in the spin-taste basis.

\begingroup
\squeezetable
\begin{table*}
\renewcommand{\arraystretch}{1.05}
\renewcommand{\tabcolsep}{1pt}
\begin{tabular}{|c|c|c|c|}
\hline
No. 
& $\Sigma_{Q}^{(*)}
: (\bf{3}_{S},\bf{3}_{S},\bf{10}_{S})_{\rm 0}$
& $\Xi'^{(*)}_{Q}
: (\bf{3}_{S},\bf{2},\bf{10}_{S})_{\rm -1}$
& $\Omega_{Q}^{(*)}
: (\bf{3}_{S},\bf{1},\bf{10}_{S})_{\rm -2}$ \\ \hline
$3$ & $\eta_{k}\epsilon\zeta_{k}ll$ 
& $\eta_{k}\epsilon\zeta_{k}ls+\eta_{k}\epsilon\zeta_{k}sl$
& $\eta_{k}\epsilon\zeta_{k}ss$ \\
 $4$ & $\eta_{4}\zeta_{4}\eta_{k}\epsilon\zeta_{k}ll$ 
& $\eta_{4}\zeta_{4}\eta_{k}\epsilon\zeta_{k}ls+
\eta_{4}\zeta_{4}\eta_{k}\epsilon\zeta_{k}sl$
& $\eta_{4}\zeta_{4}\eta_{k}\epsilon\zeta_{k}ss$ \\ 
$5$ & $l\eta_{k}D_{k}l$ 
& $l\eta_{k}D_{k}s + s\eta_{k}D_{k}l$
& $s\eta_{k}D_{k}s $ \\
$9$ & $\eta_{k}\epsilon\zeta_{k}l\eta_{l}D_{l}l$
& $\eta_{k}\epsilon\zeta_{k}l\eta_{l}D_{l}s 
+\eta_{k}\epsilon\zeta_{k}s\eta_{l}D_{l}l$ 
& $\eta_{k}\epsilon\zeta_{k}s\eta_{l}D_{l}s$ \\ 
$15$ & $\eta_{m}\zeta_{m}l\eta_{k}D_{k}\{\zeta_{l}D_{l}l\}$ 
& $\eta_{m}\zeta_{m}l\eta_{k}D_{k}\{\zeta_{l}D_{l}s\}
+ \eta_{m}\zeta_{m}s\eta_{k}D_{k}\{\zeta_{l}D_{l}l\}$
& $\eta_{m}\zeta_{m}s\eta_{k}D_{k}\{\zeta_{l}D_{l}s\}$ \\
$16$ & $\eta_{4}\zeta_{4}\eta_{m}\zeta_{m}l\eta_{k}D_{k}\{\zeta_{l}D_{l}l\}$ 
& $\eta_{4}\zeta_{4}\eta_{m}\zeta_{m}l\eta_{k}D_{k}\{\zeta_{l}D_{l}s\}
+ \eta_{4}\zeta_{4}\eta_{m}\zeta_{m}s\eta_{k}D_{k}\{\zeta_{l}D_{l}l\}$
& $\eta_{4}\zeta_{4}\eta_{m}\zeta_{m}s\eta_{k}D_{k}\{\zeta_{l}D_{l}s\}$ \\
$19$ 
& $\eta_{k}\epsilon\zeta_{k}l\eta_{1}D_{1}\{\eta_{2}D_{2}\{\eta_{3}D_{3}l\}\}$
& $\eta_{k}\epsilon\zeta_{k}l\eta_{1}D_{1}\{\eta_{2}D_{2}\{\eta_{3}D_{3}s\}\}
+ \eta_{k}\epsilon\zeta_{k}s\eta_{1}D_{1}\{\eta_{2}D_{2}\{\eta_{3}D_{3}l\}\}$
& $\eta_{k}\epsilon\zeta_{k}s\eta_{1}D_{1}\{\eta_{2}D_{2}\{\eta_{3}D_{3}s\}\}$
\\ \hline
\end{tabular}
\begin{tabular}{|c|c|c|}
\hline
No. 
& $\Sigma_{Q}^{(*)}
: (\bf{3}_{S},\bf{1}_{A},\bf{6}_{A})_{\rm 0}$ 
& $\Xi'^{(*)}_{Q}
: (\bf{3}_{S},\bf{2},\bf{6}_{A})_{\rm -1}$ \\ \hline
$6$ & $\eta_{4}\zeta_{4}l_{1}\eta_{k}D_{k}l_{2} 
- \eta_{4}\zeta_{4}l_{2}\eta_{k}D_{k}l_{1}$ 
& $\eta_{4}\zeta_{4}l\eta_{k}D_{k}s 
- \eta_{4}\zeta_{4}s\eta_{k}D_{k}l$ \\
$10$ & $\eta_{4}\zeta_{4}\eta_{k}\epsilon\zeta_{k}l_{1}\eta_{l}D_{l}l_{2}
- \eta_{4}\zeta_{4}\eta_{k}\epsilon\zeta_{k}l_{2}\eta_{l}D_{l}l_{1}$ 
& $\eta_{4}\zeta_{4}\eta_{k}\epsilon\zeta_{k}l\eta_{l}D_{l}s
- \eta_{4}\zeta_{4}\eta_{k}\epsilon\zeta_{k}s\eta_{l}D_{l}l$ \\ 
$11$ 
& $l_{1}\eta_{k}D_{k}\{\eta_{l}D_{l}l_{2}\}
- l_{2}\eta_{k}D_{k}\{\eta_{l}D_{l}l_{1}\}$
& $l\eta_{k}D_{k}\{\eta_{l}D_{l}s\}
- s\eta_{k}D_{k}\{\eta_{l}D_{l}l\}$ \\
$12$ 
& $\eta_{4}\zeta_{4}l_{1}\eta_{k}D_{k}\{\eta_{l}D_{l}l_{2}\}
- \eta_{4}\zeta_{4}l_{2}\eta_{k}D_{k}\{\eta_{l}D_{l}l_{1}\}$
& $\eta_{4}\zeta_{4}l\eta_{k}D_{k}\{\eta_{l}D_{l}s\}
- \eta_{4}\zeta_{4}s\eta_{k}D_{k}\{\eta_{l}D_{l}l\}$ \\
$20$ 
& $\eta_{4}\zeta_{4}\eta_{k}\epsilon\zeta_{k}l_{1}
\eta_{1}D_{1}\{\eta_{2}D_{2}\{\eta_{3}D_{3}l_{2}\}\}
-\eta_{4}\zeta_{4}\eta_{k}\epsilon\zeta_{k}l_{2}
\eta_{1}D_{1}\{\eta_{2}D_{2}\{\eta_{3}D_{3}l_{1}\}\}$ 
& $\eta_{4}\zeta_{4}\eta_{k}\epsilon\zeta_{k}l
\eta_{1}D_{1}\{\eta_{2}D_{2}\{\eta_{3}D_{3}s\}\}
-\eta_{4}\zeta_{4}\eta_{k}\epsilon\zeta_{k}s
\eta_{1}D_{1}\{\eta_{2}D_{2}\{\eta_{3}D_{3}l\}\}$
 \\ \hline
\end{tabular}
\begin{tabular}{|c|c|c|}
\hline
No. 
& $\Lambda_{Q} : (\bf{1}_{A},\bf{1}_{A},\bf{10}_{S})_{\rm 0}$ 
& $\Xi_{Q} : (\bf{1}_{A},\bf{2},\bf{10}_{S})_{\rm -1}$ \\ \hline
$7$ & $l_{1}\epsilon\zeta_{k}D_{k}l_{2} - l_{2}\epsilon\zeta_{k}D_{k}l_{1}$ 
& $l\epsilon\zeta_{k}D_{k}s - s\epsilon\zeta_{k}D_{k}l$ \\
$13$ 
& $l_{1}\zeta_{k}D_{k}\{\zeta_{l}D_{l}l_{2}\}
- l_{2}\zeta_{k}D_{k}\{\zeta_{l}D_{l}l_{1}\}$
& $l\zeta_{k}D_{k}\{\zeta_{l}D_{l}s\}
- s\zeta_{k}D_{k}\{\zeta_{l}D_{l}l\}$ \\
$14$ 
& $\eta_{4}\zeta_{4}l_{1}\zeta_{k}D_{k}\{\zeta_{l}D_{l}l_{2}\}
- \eta_{4}\zeta_{4}l_{2}\zeta_{k}D_{k}\{\zeta_{l}D_{l}l_{1}\}$
& $\eta_{4}\zeta_{4}l\zeta_{k}D_{k}\{\zeta_{l}D_{l}s\}
- \eta_{4}\zeta_{4}s\zeta_{k}D_{k}\{\zeta_{l}D_{l}l\}$ \\
$17$ 
& $l_{1}\eta_{1}D_{1}\{\eta_{2}D_{2}\{\eta_{3}D_{3}l_{2}\}\}
- l_{2}\eta_{1}D_{1}\{\eta_{2}D_{2}\{\eta_{3}D_{3}l_{1}\}\} $
& $l\eta_{1}D_{1}\{\eta_{2}D_{2}\{\eta_{3}D_{3}s\}\}
- s\eta_{1}D_{1}\{\eta_{2}D_{2}\{\eta_{3}D_{3}l\}\} $
\\ \hline
\end{tabular}
\begin{tabular}{|c|c|c|}
\hline
No. 
& $\Lambda_{Q} : (\bf{1}_{A},\bf{3}_{S},\bf{6}_{A})_{\rm 0}$ 
& $\Xi_{Q} : (\bf{1}_{A},\bf{2},\bf{6}_{A})_{\rm -1}$ \\ \hline
$1$ & $ll$ & $ls+sl$ \\
$2$ & $\eta_{4}\zeta_{4}ll$ & $\eta_{4}\zeta_{4}ls+\eta_{4}\zeta_{4}sl$ \\
$8$ & $\eta_{4}\zeta_{4}l\epsilon\zeta_{k}D_{k}l$ 
& $\eta_{4}\zeta_{4}l\epsilon\zeta_{k}D_{k}s
+ \eta_{4}\zeta_{4}s\epsilon\zeta_{k}D_{k}l$ \\
$18$ 
& $\eta_{4}\zeta_{4}l\eta_{1}D_{1}\{\eta_{2}D_{2}\{\eta_{3}D_{3}l\}\}$
& $\eta_{4}\zeta_{4}l\eta_{1}D_{1}\{\eta_{2}D_{2}\{\eta_{3}D_{3}s\}\}
+ \eta_{4}\zeta_{4}s\eta_{1}D_{1}\{\eta_{2}D_{2}\{\eta_{3}D_{3}l\}\}$
\\ \hline
\end{tabular}
\caption{Lattice staggered diquark operators categorized into the continuum 
irreps $(SU(2)_{S},SU(2)_{I},SU(4)_{T})_{Z}$ : 
The $l$ and $s$ denote light and strange quark, respectively.
No. indicates the operator number in Table 
\ref{diquark_operators}
and \ref{diquark_operators2}.
The summation over $\bf{x}$ and color indices are omitted.}
\label{result}
\end{table*}
\endgroup


\section{Connection between lattice and continuum irreps}

Consulting the relations between lattice $\overline{RF}$ irreps and
continuum spin irreps given in \cite{Golterman}
and assuming that the ground states of lattice irreps correspond to 
the lowest possible spin in the continuum limit,
one could make an assignment of spin and parity $J^{P}$
for each $GTS$ irrep.
See the seventh column of
Tables \ref{diquark_operators} and \ref{diquark_operators2}. 
One also see that
the combinations, 
$
C\Gamma_{S} = C\gamma_{k}, 
C\gamma_{k}\gamma_{4},
C\gamma_{4}\gamma_{5},
C\gamma_{5}, 
$
generate $upper \times upper$ 
products of the Dirac spinors 
in Dirac representation 
for each taste
and then give rise to 
$\mathcal{O}(1)$ contributions,
while the combinations,
$
C\Gamma_{S} = C, 
C\gamma_{4},
C\gamma_{k}\gamma_{l},
C\gamma_{k}\gamma_{5}, 
$
induce $upper \times lower$ products,
so that they are suppressed by $\mathcal{O}(p/E)$
in the non-relativistic limit.
See the second last column of Table \ref{diquark_operators} and
\ref{diquark_operators2}.
It is important to 
notice that only the positive parity states 
survive in the non-relativistic limit,
although every $GTS$ irrep contains
both parities. 
This is in accordance with the nature of
physical diquarks.
As for the $SU(4)_{T}$ irreps, one sees that 
the combinations,
$
\Gamma_{T}C^{-1} = \gamma_{k}C^{-1}, 
\gamma_{4}C^{-1},
\gamma_{k}\gamma_{l}C^{-1},
\gamma_{k}\gamma_{4}C^{-1},
$
are symmetric 
so that they  
altogether correspond to $\bf{10}_{S}$ of $SU(4)_{T}$,
while the anti-symmetric 
six combinations
$
\Gamma_{T}C^{-1} = C^{-1},
\gamma_{k}\gamma_{5}C^{-1},
\gamma_{4}\gamma_{5}C^{-1},
\gamma_{5}C^{-1},
$
belong to
$\bf{6}_{A}$ of $SU(4)_{T}$.
The assignments of non-relativistic $SU(2)_{S}\times SU(4)_{T}$ 
for the lattice irreps
are readily given for the $\mathcal{O}(1)$ operators. 
They are listed  in the last column of 
Table \ref{diquark_operators} and \ref{diquark_operators2}.

The final step 
is to take into account the $2+1$ flavor symmetry,
which could be done in a straightforward manner.
The decomposition of continuum spin, $2+1$ flavor and taste 
symmetry group  
into the lattice symmetry group is given by,
\begin{eqnarray}
SU(2)_{S}\times SU(2)_{I}\times SU(4)_{T}
&\supset& SU(2)_{I}\times GTS. \qquad
\end{eqnarray}
Table \ref{result} summarizes all the local-time lattice diquark operators
categorized into
each continuum irrep $(SU(2)_{S},SU(2)_{I},SU(4)_{T})_{Z}$
given in (\ref{Sigma_2})-(\ref{Xi_2}).

\section*{Acknowledgments}

We would like to thank S. Basak, C. Bernard and
C. DeTar for useful discussions and comments.
We thank J. Bailey for important comments on the physical states.
This work has been supported by U.S. Department of Energy, Grant No.
FG02-91ER 40661.

\end{document}